# Semi-supervised Stacked Label Consistent Autoencoder for Reconstruction and Analysis of Biomedical Signals

Anupriya Gogna, Angshul Majumdar, *Senior Member, IEEE* and Rabab Ward, *Fellow, IEEE*

*Abstract*— **Objective:** In this work we propose an autoencoder based framework for simultaneous reconstruction and classification of biomedical signals. Previously these two tasks – reconstruction and classification were treated as separate problems. This is the first work to propose a combined framework to address the issue in a holistic fashion. **Methods:** Reconstruction techniques for biomedical signals for tele-monitoring are largely based on compressed sensing (CS) based method; these are 'designed' techniques where the reconstruction formulation is based on some 'assumption' regarding the signal. In this work, we propose a new paradigm for reconstruction – we 'learn' to reconstruct. An autoencoder can be trained for the same. But since the final goal is to analyze / classify the signal we learn a linear classification map inside the autoencoder. The ensuing optimization problem is solved using the Split Bregman technique. **Results:** Experiments have been carried out on reconstruction and classification of ECG (arrhythmia classification) and EEG (seizure classification) signals. **Conclusion:** Our proposed tool is capable of operating in a semi-supervised fashion. We show that our proposed method is better and more than an order magnitude faster in reconstruction than CS based methods; it is capable of real-time operation. Our method is also better than recently proposed classification methods. **Significance:** This is the first work offering an alternative to CS based reconstruction. It also shows that representation learning can yield better results than hand-crafted features for signal analysis.

## I. Introduction

IN recent times tele-monitoring via body area networks (BAN) has received considerable interest. Here the goal is to acquire biomedical signals like ECG, EEG, PPG etc. and transmit it wirelessly to some base station for manual or automated analysis. Such a system, when developed will benefit both developed countries and developing nations. It will help health monitoring of the elderly and the differently abled. This can also be employed for mass scale data monitoring of subjects in developing nations large portions of the country do not have access to proper medical facilities.

One of the biggest challenges in telemonitoring is to develop an energy efficient BAN. At the sensor nodes there are three energy sinks – communication, sensing and processing; communication requires the largest amount of energy followed by sensing. The energy required for processing is relatively small compared to the other two. The standard technique is to compress the signal prior to transmission. However such transform coding based techniques are computationally expensive for a sensor node. In recent times, most studies use compressed sensing (CS) instead [1-6].

CS based solutions project a portion of the acquired signal (say 1 second) onto a random matrix (Gaussian, sparse binary, binomial) such that the size of the projected data is smaller than the length of the 1 second samples. Since it only requires a matrix-vector product, it is computationally cheap. There are several studies that propose energy and computationally efficient hardware for the same [6-10]. The compressed data is wirelessly transmitted to a base station where it is reconstructed using CS techniques for further analysis and monitoring. There can be several variants of the basic CS technique for reconstruction; [1], [3] and [6] use variants of sparse Bayesian learning [11]. [12-16] use the more standard CS approach for recovery; in [17-21] the signals were reconstructed using inter-channel and intra-channel correlations.

All these studies concentrated on the reconstruction of the signal with the assumption that the signals will be analyzed retrospectively. Such offline analysis can only be done for non-critical applications like emotion assessment [22, 23]. However in most applications of biomedical signal analysis this is not the case; for example in seizure detection [24, 25] the analysis / monitoring should be in real-time.

In such time-critical systems, CS will fail. CS requires solving an optimization problem iteratively. The time required to solve the reconstruction problem on a standard PC is much larger than the time duration of the signal. In this work we reconstruct based on a deep learning approach. It is based on the extensions of the seminal paper by Hinton [26] which showed autoencoders can 'learn' to compress the information content of signals. Later studies [27, 28] showed that autoencoders can be used for denoising. In this work we go a step further and show that autoencoders can be used for solving inverse

This work was supported by Qatar National Research Fund (QNRF) no. NPRP 09 - 310 - 1 – 058.
A. Gogna is with Indraprasatha Institute of Information Technology, Delhi, India. *A. Majumdar is with Indraprasatha Institute of Information Technology.

Rabab Ward is with University of British Columbia, Vancouver, Canada (correspondence e-mail: angshul@iiitd.ac.in).

problems like reconstruction. Unlike CS based techniques our proposed method requires only a few matrix vector-products and hence can operate in real-time.

In biomedical signal processing applications the end goal is not signal reconstruction, but signal classification. Estimation emotional state, detecting seizures can all be translated to classification problems. Usually hand-crafted features [25], [29-31] or statistical features [24], [32] and [33] are extracted from the signals and standard classification tools like neural networks and support vector machines are employed. In recent times, systematic studies have shown that image analysis (for example biometrics [34] and speech processing (slightly old short review on speech recognition by deep learning [35]), automatically generated deeply learnt features yield far superior results compared to hand-crafted or statistical features. The main challenge of deep learning is the requirement for large amount of training data. This is difficult to find in biomedical signal analysis problems; probably that is the reason behind the relative sparsity of papers in this area.

In the tele-monitoring scenario there are two tasks - signal reconstruction and signal analysis. In this work we propose a combined solution to the two. The standard autoencoder is self-supervised, i.e. the input and the output are the same; they are unsupervised in the sense that they do not require any training data. In this work we learn a semi-supervised autoencoder. Along with reconstruction, it will also learn a linear map to the class labels when class information is available; for signals having no class information it will just learn to reconstruct. Such an autoencoder serves the dual purpose. It can be used for automated signal analysis; and also for manual monitoring of the reconstructed signals – both in real-time.

The rest of the paper is organized into several sections. The following literature review section discusses prior CS based reconstruction techniques and the basics of autoencoder. The proposed architecture and its implementation is described in detain in section 3. Thorough experimental evaluation is carried out in section 4. The conclusions of this work is discussed in section 5.

## II. Literature Review

### A. Compressed Sensing based Reconstruction

In this work we will talk mostly about EEG reconstruction since majority of the work has been on this area. But the techniques discussed are generic enough to be applied to any other biomedical signal.

One of the earliest works that applied CS for EEG signal compression and transmission is [12]. It projected the EEG signal onto an i.i.d Gaussian basis for compression and used CS to recover the EEG signal by exploiting the signal's sparsity in the Gabor domain. The compression can be expressed as –

$$b = \Phi z \quad (1)$$

where z is the EEG signal, $\Phi$ is the projection / compression matrix and b is the compressed data. It was assumed that the data is sparse in some domain $\Psi$ so that the sparse coefficients ($\alpha$) could be recovered by $l_1$-norm minimization.

$$\min_{\alpha} \|\alpha\|_1 \text{ subject to } b = \Phi \Psi \alpha \quad (2)$$

In [15] the authors showed that a better way to recover the signals is to use an analysis prior formulation instead of (2).

$$\min_{z} \|\Psi z\|_1 \text{ subject to } b = \Phi z \quad (3)$$

EEG is always acquired by multiple channels; ECG too is acquired from several channels. The aforementioned techniques reconstruct one channel at a time. The possibility of exploiting inter-channel correlation in order to improve EEG signal reconstruction was mentioned in [12], but there was no concrete formulation. This problem is partially addressed in [16]. They do not explicitly model the inter-channel correlation, but frame a joint reconstruction problem where the signals from all the channels are reconstructed simultaneously.

Let 'i' denote the channel number, then the compression for this channel can be represented as –

$$b_i = \Phi z_i \quad (4)$$

This can be organized as follows,

$$\begin{bmatrix} b_1 \\ \dots \\ b_C \end{bmatrix} = \begin{bmatrix} \Phi & \dots & 0 \\ \dots & \dots & \dots \\ 0 & \dots & \Phi \end{bmatrix} \begin{bmatrix} z_1 \\ \dots \\ z_C \end{bmatrix} \quad (5)$$

The concatenated solution will be sparse in wavelet domain; this sparsity of the signals from all channels is exploited in [16]. At a first glance this looks like a joint reconstruction problem, but a closer look reveals that this is actually as good as channel by channel reconstruction; this formulation (5) does not exploit any structure across the channels. Other studies assume a block structure of the EEG signals [1], [3], [6] and [11] in a transform domain (DCT or wavelet).

A recent work proposed CS techniques for EEG signal compression and transmission, but instead of sending the raw EEG signals it subtracted the mean from all the signals thereby reducing the number of bits to be transmitted [13]. However, such a scheme will not be energy efficient, since in order to compute the mean EEG signal, the nodes need to communicate with each other – and such communication consumes considerable energy.

In another work [14], it was shown that for certain specific tasks like seizure detection, instead of sending the full signal, it is possible to send some distinct features which can be further analyzed at the base station for possible risks. Such a technique requires more computation than standard CS techniques, but the number of features to be transmitted are very few. Unfortunately such a technique cannot be generalized for other applications.

Biomedical signals are inherently correlated. Prior studies hinted at using the inter-channel correlation but did not propose any formulation to exploit this information. In [18] the common inter-channel sparsity pattern was exploited to capture the correlation. Instead of (5), the organization is –

$$[b_1 | \dots | b_C] = \Phi [z_1 | \dots | z_C] \quad (6)$$

The signals from different channels being correlated will share a common sparsity pattern in the transform domain. Thus the matrix $\Psi[z_1|...|z_C]$ will be row-sparse. Hence can be recovered by $l_{2,1}$-minimization.

$$\min_Z \|Z\|_{2,1} \text{ subject to } B = \Phi Z \quad (7)$$

where $B = [b_1|...|b_C]$ and $Z = [z_1|...|z_C]$

Here the $l_{2,1}$-norm is defined as the sum of the $l_2$-norm of the rows. The outer $l_1$-norm (summation) promotes sparsity in the selection of rows. The inner $l_2$-norm promotes a dense solution in the selected row.

In [17], a separate take on correlation is proposed. The authors argued that if the signals are correlated they will be linearly dependent; therefore when stacked as columns will form a low-rank matrix, i.e. Z (7) will be low-rank. This property was exploited in [17]; a matrix completion based formulation was utilized to recover the signal ensemble.

$$\min_Z \|Z\|_{NN} \text{ subject to } B = \Phi Z \quad (8)$$

The nuclear norm (defined as the sum of nuclear values) is the closest convex surrogate of rank.

Some recent studies [19, 20] exploited the Blind Compressive Sensing (BCS) formulation; here instead of using a fixed sparsifying basis like wavelet / Gabor, it is learnt from the data.

B. *Autoencoder*

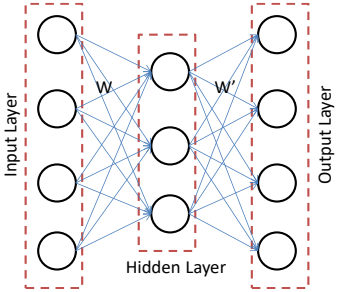

Fig. 1. Single Layer Autoencoder

An auto encoder (as seen in Fig. 1) consists of two parts – the encoder maps the input to a latent space, and the decoder maps the latent representation to the data. For a given input vector (including the bias term) $x$, the latent space is expressed as:

$$h = Wx \quad (9)$$

Here the rows of $W$ are the link weights from all the input nodes to the corresponding latent node. The mapping can be linear (9), but in most cases it is non-linear (sigmoid, tanh etc.):

$$h = \phi(Wx) \quad (10)$$

The decoder portion reverse maps the latent features to the data space.

$$x = W'\phi(Wx) \quad (11)$$

Since the data space is assumed to be the space of real numbers, there is no sigmoidal function here.

During training, the problem is to learn the encoding and decoding weights – $W$ and $W'$. This is achieved by minimizing the Euclidean cost:

$$\arg\min_{W,W'} \|X - W'\phi(WX)\|_F^2 \quad (12)$$

Here $X = [x_1|...|x_N]$ consists all the training sampled stacked as columns. The problem (12) is clearly non-convex. However, it is solved easily by gradient descent techniques since the sigmoid function is smooth and continuously differentiable.

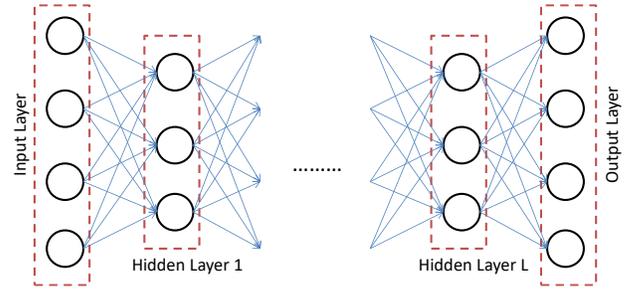

Fig. 2. Stacked Autoencoder

There are several extensions to the basic autoencoder architecture. Stacked / Deep autoencoders [26], [27] have multiple hidden layers (see Fig. 2). The corresponding cost function is expressed as follows:

$$\arg\min_{W_1...W_{L-1},W'_1...W'_L} \|X - g \circ \quad (13)$$

where $g = W_1'\phi(W_2'...W_L'(f(X)))$ and

$f = \phi(W_L\phi(W_{L-1}...\phi(W_1X)))$ .

Solving the complete problem (13) is computationally challenging. The weights are usually learned in a greedy fashion – one layer at a time [36], [37].

Stacked denoising autoencoders (SDAE) [27] are a variant of the basic autoencoder where the input consists of noisy samples and the output consists of clean samples. Here the encoder and decoder are learnt to denoise noisy input samples. The learned features appear to be more robust when learnt by SDAE.

In a recent work a marginalized denoising autoencoder was proposed [38], which does not have any intermediate nodes but learns the mapping from the input to the output. This formulation is convex (unlike regular autoencoders); the trick here is to marginalize over all possible noisy samples so that the dataset need not be augmented like SDAE. Such an autoencoder was used for domain adaptation.

Another variation for the basic autoencoder is to regularize it, i.e.

$$\arg\min_{(W)s} \|X - g \circ \quad + R(W,X) \quad (14)$$

The regularization can be a simple Tikhonov regularization – however that is not used in practice. It can be a sparsity promoting term [28], [39] or a weight decay term (Frobenius norm of the Jacobian) as used in the contractive autoencoder [40]. The regularization term is usually chosen so that they are differentiable and hence minimized using gradient descent techniques.

In a recent work [41] a group-sparse autoencoder is proposed. Here the regularization term is an $l_{2,1}$-norm on each class. The idea is that by enforcing a similar sparsity signature in each class one can enforce some level of supervision in autoencoding. This work is different from ours.

## III. PROPOSED APPROACH

In this section we discuss our proposed label consistent autoencoder. This will be divided into two sub-sections. In the first one will discuss why it is possible to recover the signals using autoencoders. In the second sub-section we add the label consistency term and complete the formulation.

### A. Reconstruction

We are interested in solving a linear inverse problem of the form: $y = Ax$. For a determined or an over-determined system the solution is linear. Even for an under-determined system the minimum energy solution is linear. However, a compressed sensing solution is non-linear. First we discuss the reason behind this non-linearity.

Ideally to obtain a sparse solution one would like to solve the $l_0$-minimization problem, but as is well known, this is NP hard.

$$\min_x \|x\|_0 \text{ subject to } y = Ax \qquad (15)$$

One way to solve (15) is to employ greedy techniques such as orthogonal matching pursuit (OMP) [41]. This is a greedy approach which detects one support (non-zero position in x) at a time and estimates its value. The full algorithm is given by,

**Algorithm OMP**

---
**Input**: y, A, k (support)
**Initialize**: $r = y, \Omega = \emptyset$ (support set)
**Repeat for k iterations**
  Compute Correlation: $c = abs(A^T r)$
  Detect Support: $l = \arg\max_i c_i$
  Update Support: $\Omega = \Omega \cup l$
  Estimate values at support $\Omega$: $x_\Omega = \min_x \|y - A_\Omega x_\Omega\|_2^2$
  Compute residual: $r = y - A_\Omega x_\Omega$
**End**
---

Here the subscript $\Omega$ means that only those columns in A indexed in $\Omega$ are selected. After the iterations are over, we get a solution with the values at some non-zero positions. To get the full x, one needs to fill in the other positions with 0 values.

OMP is a non-linear operation. In every iteration, one needs to compute the 'max' during the support detection stage – this is a highly non-linear operation. Extensions of OMP like StOMP [42], or CoSamp [43] are also non-linear. StOMP requires a thresholding operation; CoSamp requires a sorting – both are non-linear operations.

So far we have talked about greedy algorithms for sparse recovery. The more popular technique is to relax the NP hard $l_0$-norm to its closest convex surrogate the $l_1$-norm. This enjoys stronger theoretical guarantees. In practice the solution is obtained via:

$$\min_x \|y - Ax\|_2^2 + \lambda \|x\|_1 \qquad (16)$$

Consider the simplest technique to solve (16) – Iterative Soft Thresholding Algorithm (ISTA) [44]. Every iteration (say k) consists of two steps. The first step is the Landweber Iteration (17) and the second step is the soft thresholding (18).

$$b = x_{k-1} + \sigma A^T (y - Ax_{k-1}) \qquad (17)$$

$$x_k = sign(b) \max\left(0, |b| - \frac{\lambda\sigma}{2}\right) \qquad (18)$$

Where the step-size σ is inverse of the maximum Eigenvalue of $A^T A$.

The first step (17) is a simple gradient descent step – it is a linear operation. But the second step involves thresholding and is hence a non-linear operation.

To summarize, all CS recovery techniques are non-linear inversion operations.

CS has been used extensively in the past decade in a variety of signal processing applications, ranging from biomedical signal reconstruction to medical imaging, seismic imaging and astronomy to name a few. However CS cannot be used for real-time reconstruction, since the solution is iterative and hence time consuming. In our problem, the reconstruction should be real-time; this precludes use of such sophisticated inversion techniques. Instead of designing the inversion techniques (discussed in section II) we will 'learn' to reconstruct.

It is well known that Neural Networks act as universal function approximators. Given enough training data the non-linear activation functions learn to represent arbitrary functions; this was proven by Cybenko [46] and Hornik [47]. A more fundamental work on this topic dates back to Kolmogorov [48] where he showed that a continuous function of many variables can be approximated by a superposition of continuous functions of one variable. We make use of this universal functional approximation property of neural networks to 'learn' a CS like inversion operation with autoencoders.

The basic idea is simple, a poor man's inverse of a linear system is obtained by:

$$x' = A^T y = A^T Ax \qquad (19)$$

This x' is a noisy version of the actual solution x. In CS, the noise is progressively removed by soft-thresholding [49]. It has been shown that autoencoders can also be used for the denoising [28]. Basically one prepares a large number of noisy versions of the signal and input them to the autoencoder; at the output are the corresponding noisy versions. The autoencoder 'learns' a mapping from the noisy input to the clean output. When a new noisy signal is later applied to the input, a clean version of it is obtained at the output. In [28] it was claimed that the simple denoising autoencoder can yield decent denoising results – sometimes even at par with dictionary learning based techniques.

In this work the 'noisy' signals are the obtained from poor man's inverse (x'); these are input to the autoencoder for training. The corresponding clean signals are at the output. During training, the autoencoder learns to 'clean' the signal. The training time can be large, but during actual operation one only needs two (for a single layer autoencoder) matrix vector products; therefore it is super-fast.

## B. Combined Classification and Reconstruction

Whatever we have discussed so far can be done using a classical autoencoder. But such an autoencoder cannot classify. One can follow the usual deep learning approach where after learning the autoencoder, the decoder is removed and a softmax layer is attached and the full architecture is fine-tuned for classification. Such a deep neural network would classify but could not reconstruct. As mentioned before, our problem demands both – reconstruction as well as classification. Some of the tasks may be automated (and would not require reconstruction) but many others would require manual monitoring. This would require a novel autoencoder that can simultaneously reconstruct and classify.

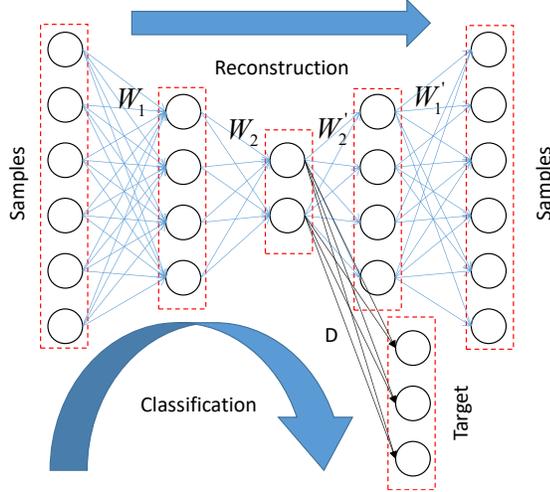

Fig. 3. Proposed Label Consistent Autoencoder

Our proposed architecture is shown in Fig. 3. It is a two layer stacked autoencoder. For all deep learning tasks, the features from the deepest layer are used; therefore we propose to learn a linear map from the innermost layer to the targets; this constitutes the label consistency criterion. This idea has been used in the past for discriminative DBM [50] and label consistent dictionary learning [51]. The mathematical expression is given by,

$$\min_{W_1',W_2',W_2,W_1,D} \|X - W_1'\phi(W_2'\phi(W_2\phi(W_1 X)))\|_F^2 \\ + \lambda \|T - D\phi(W_2\phi(W_1 X))\|_F^2 \quad (20)$$

Here X is the training samples, T the targets and D the linear map. It is not possible to learn this architecture using off-the-shelf backpropagation techniques. This is because there are two outputs, therefore there is no unique unambiguous way to backpropagate the errors. We will solve it using the Split Bregman technique. But before, getting into the solution, we need to incorporate semi-supervision, i.e. not all the training samples will have a label. This leads to:

$$\min_{W_1',W_2',W_2,W_1,D} \|X - W_1'\phi(W_2'\phi(W_2\phi(W_1 X)))\|_F^2 \\ + \lambda \|T - D\phi(W_2\phi(W_1 X_S))\|_F^2 \quad (21)$$

We assume the training data to be $X = [X_U | X_S]$, where the subscripts denote Unsupervised or Supervised.

## C. Derivation

In this work we solve (21) by a Bregman type variable splitting [52]. In the first step we substitute

$$Z^1 = \phi(W_2'\phi(W_2\phi(W_1 X))) \quad (22)$$

The proxy variable has two parts – unsupervised and supervised, i.e. $Z^1 = [Z_U^1 | Z_S^1]$. This allows us to express (21) as follows:

$$\min_{W_1',W_2',W_2,W_1,D,Z^1} \|X - W_1'Z^1\|_F^2 + \lambda \|T - D\phi(W_2\phi(W_1 X))\|_F^2 \\ s.t. \ Z^1 = \phi(W_2'\phi(W_2\phi(W_1 X))) \quad (23)$$

One can incorporate the proxy and variables by a Lagrangian, but the exact Lagrangian would enforce equality between the two in every iteration. This is not required; for practical purposes we only need the proxy and the variables to converge at the solution. Therefore one can relax the Lagrangian to the augmented Lagrangian instead:

$$\min_{W_1',W_2',W_2,W_1,D,Z^1} \|X - W_1'Z^1\|_F^2 + \lambda \|T - D\phi(W_2\phi(W_1 X))\|_F^2 \\ + \mu_1 \|Z^1 - \phi(W_2'\phi(W_2\phi(W_1 X)))\|_F^2 \quad (24)$$

In the Augmented Lagrangian formulation, one starts with a small value of μ – this relaxes the equality constraint. For each value of μ, (24) is solved and then then value of μ is increased to enforce equality progressively. As one can see this is not an elegant approach; increasing the value of μ is at best heuristic. The most elegant solution is to incorporate a Bregman relaxation variable $B_1$, this automatically adjusts for the equality constraint since it is updated. One does not need to tune the values of μ. The Split Bregman formulation is:

$$\min_{W_1',W_2',W_2,W_1,D,Z^1} \|X - W_1'Z^1\|_F^2 + \lambda \|T - D\phi(W_2\phi(W_1 X))\|_F^2 \\ + \mu_1 \|Z^1 - \phi(W_2'\phi(W_2\phi(W_1 X))) - B_1\|_F^2 \quad (25)$$

We apply the Split Bregman technique on the substitution $Z^2 = \phi(W_2\phi(W_1 X))$, leading to:

$$\min_{W_1',W_2',W_2,W_1,D,Z^1,Z^2} \|X - W_1'Z^1\|_F^2 + \lambda \|T - DZ_S^2\|_F^2 \\ + \mu_1 \|Z^1 - \phi(W_2'Z^2) - B_1\|_F^2 \quad (26) \\ + \mu_2 \|Z^2 - \phi(W_2\phi(W_1 X)) - B_2\|_F^2$$

As in $Z^1$, $Z^2$ has two parts – Supervised (denoted by subscript S) and Unsupervised (denoted by subscript U). In the third level, we substitute $Z = \phi(W_1 X)$. This leads to the final formulation:

$$\min_{W_1',W_2',W_2,W_1,D,Z^1,Z^2,Z} \|X - W_1'Z^1\|_F^2 + \lambda \|T - DZ_S^2\|_F^2 \\ + \mu_1 \|Z^1 - \phi(W_2'Z^2) - B_1\|_F^2 \quad (27) \\ + \mu_2 \|Z^2 - \phi(W_2 Z) - B_2\|_F^2 + \mu \|Z - \phi(W_1 X) - B\|_F^2$$

Even though not exactly separable, (27) can be segregated into a number of sub-problems:

$$P1: \min_{W_1'} \|X - W_1'Z^1\|_F^2$$

$$P2: \min_{W_2'} \|Z^1 - \phi(W_2'Z^2) - B_1\|_F^2 \equiv \|\phi^{-1}(Z^1 - B_1) - W_2'Z^2\|_F^2$$

$$P3: \min_{W_2} \|Z^2 - \phi(W_2 Z) - B_2\|_F^2 \equiv \|\phi^{-1}(Z^2 - B_2) - W_2 Z\|_F^2$$

$$P3: \min_{W_1} \|Z - \phi(W_1 X) - B\|_F^2 \equiv \|\phi^{-1}(Z - B) - W_1 X\|_F^2$$

$$P4: \arg\min_{Z^1} \|X - W_1'Z^1\|_F^2 + \mu_1 \|Z^1 - \phi(W_2'Z^2) - B_1\|_F^2$$

$$P5: \min_{Z^2} \lambda \|T - DZ_S^2\|_F^2 + \mu_1 \|Z^1 - \phi(W_2'Z^2) - B_1\|_F^2$$
$$+ \mu_2 \|Z^2 - \phi(W_2 Z) - B_2\|_F^2$$

$$P6: \min_{Z} \mu_2 \|Z^2 - \phi(W_2 Z) - B_2\|_F^2 + \mu \|Z - \phi(W_1 X) - B\|_F^2$$

$$P7: \min_{D} \|T - DZ_S^2\|_F^2$$

Sub-problem P1 and P7 are a simple least squares problems having a closed form solutions. Sub-problems P1-P3 can be recast as linear least squares problems (shown above) and hence can be solved analytically as well. Sub-problem P4 can be re-arranged as follows,

$$\min_{Z^1} \left\| \begin{pmatrix} X \\ \sqrt{\mu}\left(\phi(W_2'Z^2) + B_1\right) \end{pmatrix} - \begin{pmatrix} W_1' \\ \sqrt{\mu} I \end{pmatrix} Z^1 \right\|_F^2 \quad (28)$$

This turns out to be a simple least squares problem as well. Similarly one can recast P6 as a least squares problem in the following manner.

$$\min_{Z} \mu_2 \left\| \begin{pmatrix} \phi^{-1}(Z^2 + B_2) \\ \sqrt{\mu}\left(\phi(W_1 X) + B\right) \end{pmatrix} - \begin{pmatrix} W_2 \\ \sqrt{\mu} I \end{pmatrix} Z \right\|_F^2 \quad (29)$$

Sub-problem P5 can be expressed in two parts:

$$P5: \arg\min_{Z_U^2, Z_S^2} \lambda \|T - DZ_S^2\|_F^2 + \mu_1 \|Z^1 - \phi(W_2'[Z_U^2 | Z_S^2]) - B_1\|_F^2$$
$$+ \mu_2 \|[Z_U^2 | Z_S^2] - \phi(W_2 Z) - B_2\|_F^2$$

The variables $Z_U^2, Z_S^2$ are separable. Hence P5 can be segregated as follows:

$$\min_{Z_U^2} \mu_1 \|Z^1 - \phi(W_2'Z_U^2) - B_1\|_F^2 + \mu_2 \|Z_U^2 - \phi(W_2 Z) - B_2\|_F^2 \quad (30)$$

$$\min_{Z_S^2} \lambda \|T - DZ_S^2\|_F^2 + \mu_1 \|Z^1 - \phi(W_2'Z_S^2) - B_1\|_F^2$$
$$+ \mu_2 \|Z_S^2 - \phi(W_2 Z) - B_2\|_F^2 \quad (31)$$

As we have been doing so far, we can recast (30) and (31) as least squares problems (32) and (33) respectively.

$$\min_{Z_U^2} \left\| \begin{pmatrix} \sqrt{\mu_1}\phi^{-1}(Z^1 + B_1) \\ \sqrt{\mu_2}\left(\phi(W_2 Z) + B_2\right) \end{pmatrix} - \begin{pmatrix} \sqrt{\mu_1} W_2' \\ \sqrt{\mu_2} I \end{pmatrix} Z_U^2 \right\|_F^2 \quad (32)$$

$$\min_{Z_S^2} \left\| \begin{pmatrix} \sqrt{\mu_1}\phi^{-1}(Z^1 + B_1) \\ \sqrt{\mu_2}\left(\phi(W_2 Z) + B_2\right) \\ \sqrt{\lambda} T \end{pmatrix} - \begin{pmatrix} \sqrt{\mu_1} W_2' \\ \sqrt{\mu_2} I \\ \sqrt{\lambda} D \end{pmatrix} Z_S^2 \right\|_F^2 \quad (33)$$

The last part is to update the Bregman relaxation variables. This accounts for the automatic adjustments between the variables and their proxies at convergence. The relaxation variables are updated using simple gradient descent.

$$B_1 \leftarrow Z^1 - \phi(W_2'Z^2) - B_1$$
$$B_2 \leftarrow Z^2 - \phi(W_2 Z) - B_2$$
$$B \leftarrow Z - \phi(W_1 X) - B$$

There are two stopping criteria. Iterations continue till the objective function reaches some local minima, i.e. there is no significant change in successive iterations. Or, they continue for a fixed number of iterations. Our algorithm requires specifying several hyper-parameters. We found that they are very robust to a wide range of values; in this work we put $\mu_1 = \mu_2 = \mu = 0.01$

## IV. Experimental Evaluation

### A. ECG arrhythmia classification and reconstruction

In this study, five types of beat classes of arrhythmia as recommended by Association for Advancement of Medical Instrumentation (AAMI) were analyzed from ECG signals namely: non-ectopic beats, supra-ventricular ectopic beats, ventricular ectopic beats, fusion betas and unclassifiable and paced beats. The classification experiments are carried out on the MIT-BIH Arrhythmia dataset from www.physionet.org. First we carry out reconstruction and classification using the aforesaid database. This is a fully supervised problem, i.e. all the samples have class labels.

It is well known in deep learning that 'more the merrier'. However in real-life supervised samples are few; but it is easy to have a large number of unsupervised samples. Therefore in the second set of experiments we augment the aforesaid (supervised) dataset MIT-BIH with the European ST-T dataset (cardiac ischemia) dataset from www.physionet.org. No class information from the second dataset is used; it is only used for semi-supervised learning.

The MIT-BIH Arrhythmia database contains 48 half hour recordings of two channel ambulatory ECG, obtained from 47 subjects in the year 1975 and 1979 by the Beth-Israel Hospital Arrhythmia Laboratory at Boston. Twenty-four hour ambulatory ECG recordings were collected from a mixed population of size 4000 having inpatients (around 60%) and outpatients (around 40%). The recordings were digitized at 360 samples per second per channel with 11-bit resolution over a 10 mV range. Two or more cardiologists independently annotated each record; consensus was made to obtain the computer-readable reference annotations for each beat.

European society of cardiology has provided a standard ST-T database consisting of 90 annotated samples of ambulatory ECG recordings from 79 subjects having myocardial ischemia disease. The subjects were 70 men aged from 30 to 84 years, and some women aged from 55 to 71 years. Additional selection criteria were established in order to obtain a representative selection of ECG abnormalities in the database, including baseline ST segment displacement resulting from conditions such as hypertension, ventricular dyskinesia, and effects of medication. Each record is of 2 h duration and contains two signals. Each is sampled at 250 samples per second with 12-bit resolution over a nominal 20 mV input range.

As a pre-processing step the MIT-BIH dataset is down-sampled to 250 Hz from its native 360 Hz; this is ensure parity between the two datasets. Both the datasets are normalized. The quantization level remains as it is. The MIT-BIH protocol is converted to the AAMI / ANSI standard. This leads to 5 classes - Non ectopic beat (N), Supra-ventricular ectopic beats (S), Ventricular ectopic beats (V), Fusion beat (F) and Unknown beat (Q). Owing to the relative sparsity of samples in the F and Q classs they are merged with V; this is following the AAMI2 protocol proposed in [53].

We train our proposed label-consistent autoencoder with one second (250 points) length samples. The outer hidden layer has 125 nodes and the inner hidden layer has 63 nodes. The class of the entire duration is assigned to the sample during training. During testing, the test ECG sequence is broken down into one second samples ($X_{test}$) and passed through the trained model. The target for this is obtained by $T_{test} = D\phi(W_2\phi(W_1 X_S))$. Practically $T_{test}$ will not have ones and zeroes, they would be real numbers in between. We take the row averages of T and assign the class of $X_{test}$ to the class having the maximum value in the corresponding row.

For the experimental protocol we follow [54]; this is repeatable protocol. The division into test set and training set is shown in Table I. The record number (#) of the patient used for training are – 101,114,122,207,223,106,115,124,208,230,108, 116,201,209,109,118,203,215,112,119,205,220; for testing are – 100,117,210,221,233,103,121,212,222,234,105,123,213,228, 111,200,214,231,113,202,219,232.

TABLE IV
TRAIN AND TEST SET DETAILS

| Dataset | N | S | V | F | Q | Total | # Rec |
|---|---|---|---|---|---|---|---|
| Train | 45844 | 943 | 3788 | 415 | 8 | 50998 | 22 |
| Test | 44238 | 1836 | 3221 | 388 | 7 | 49690 | 22 |
| Total | 90082 | 2779 | 7009 | 803 | 15 | 100688 | 44 |

In this work we emulate a health-monitoring scenario. We assume that the ECG signals are acquired and are compressed by projecting them onto a lower dimension by a sparse binary matrix [1]. The compressed data is sent to a base station for reconstruction and analysis. As a benchmark for reconstruction we use the Block Sparse Bayesian Learning (BSBL) algorithm [11] with wavelet as the sparsifying transform; this has been used in extensively in the past for reconstructing compressively sampled biological signals [1], [3] and [6]. The reconstructed signal is then used for classification. Here we compare with two recent techniques optimum-path forest (OPF), support vector machine (SVM) [54], Probabilistic Neural Network (PNN) [55] and Extreme Learning Machine (ELM) [56]; all of them use hand-crafted features. The best results are obtained from the feature extraction technique proposed in [57]; hence we use the same for our comparison. In the aforesaid references, detailed comparison have been done with other techniques and these were shown to yield the best results; hence we compare with these studies.

As mentioned before, two sets of experiments have been carried out. In the first set only the MIT-BIH database has been used. In the second set, the database has been augmented with unsupervised samples from the European ST-T database. The results for reconstruction are shown in Table II and those from classification are shown in Table III. For classification, reconstructed signals from 50% compression have been used.

For reconstruction, Normalized Mean Squared Error is the error metric.

$$NMSE = \frac{\|groundtruth - reconstructed\|_2}{\|groundtruth\|_2}$$

We report the mean reconstruction error and the deviations. Classification Accuracy (Acc.) is the most important measure for performance; but it is a standard practice to report sensitivity (Sens.) and specificity (Spec.); the standard definitions apply for all the metrics.

TABLE II
ECG RECONSTRUCTION RESULTS

| Technique | 50% Compression | 25% Compression |
|---|---|---|
| BSBL | 0.121±0.056 | 0.262±0.114 |
| Prop. | 0.140±0.014 | 0.190±0.026 |
| Prop. Aug. | **0.089±0.006** | **0.122±0.018** |

TABLE III
ECG CLASSIFICATION RESULTS (AAMI2 PROTOCOL)

| Classifier | Acc. | F | | S | | V | |
|---|---|---|---|---|---|---|---|
| | | Sens. | Spec. | Sens. | Spec. | Sens. | Spec. |
| OPF | 86.5 | 91.2 | **56.8** | 11.0 | 97.4 | 62.4 | 90.8 |
| SVM | 90.1 | **98.8** | 31.9 | 0 | 97.6 | 41.7 | 95.4 |
| PNN | 93.8 | 94.6 | 55.3 | 15.9 | 97.0 | 48.9 | 89.6 |
| ELM | 89.2 | 95.6 | 39.8 | 0 | 97.0 | 50.2 | 95.2 |
| Prop. | 92.0 | 96.8 | 54.5 | 13.6 | **100** | 48.6 | 93.6 |
| Prop. Aug. | **96.9** | **98.8** | 56.6 | **19.0** | **100** | **65.2** | **96.2** |

Prop. = MIT-BIH; Prop. Aug = MIT-BIH + European ST-T

In Table II, the results are shown for the MIT-BIH database only. Even though we augment the dataset for our technique, we do not report the results for European ST-T; this is to keep all the results in sync. We can observe that the proposed technique improves with additional data and can yield results even better than sophisticated compressed sensing techniques. The reconstruction time required by the BSBL algorithm (takes ~ 12 seconds) is about 40 times more than our proposed autoencoder (takes ~ 0.3 seconds to reconstruct signals of 1 second durations) based approach. Therefore, not only do we recover the signal more accurately, we are faster than required for real-time operation.

In classification we see that our proposed technique (even without augmentation) yields competitive results. It is among the top two results. But with augmentation, the results improve even more. We always perform the best in terms of accuracy. For a few isolated cases, our specificity and sensitivity are marginally low. One should note that, the results in Table III cannot be directly compared with [54]; this is because in the prior work the groundruth signal is used whereas in the current work the reconstructed signal is used. Therefore there is bound to be some fall in accuracy.

## B. EEG classification and reconstruction

A publicly available EEG dataset, made available by the University of Bonn [58] is used in this work. The EEG database consists of five sets (A–E). Each set contains 100 single-channel EEG segments, each with a duration of 23.6 s. Sets A and B have been recorded using the standard international 10–20 system for surface EEG recording. Five healthy volunteers were participated in these tests with eyes open (A) and eyes closed (B). For sets C, D, and E, five epileptic patients were selected for presurgical evaluation of epilepsy by using intracranial electrodes. Depth electrodes were implanted symmetrically to record EEG from the epileptogenic zone (D) and from hippocampal formation of the opposite hemisphere of the brain (C). Segments of set E were taken from contacts of all electrodes. In sets C and D, segments contain interictal intervals while seizure activities occur in set E. Each epoch was sampled at 173.61 Hz resulting in a total of 4096 samples.

Most prior studies like [24], [59-61] convert it to a binary classification problem – seizure vs non-seizure. In this work we classify all the 5 classes A to E as defined in [62]. We compare our proposed technique with empirical mode decomposition (EMD) [24] – using SVM, Rational Discrete Short-Time Fourier (DTSTF) transform [59] – using Neural Network and Linear Prediction Error (LPE) [62] – using simple thresholding. For our proposed method, the number of nodes in the outer layer is 1024 and in the inner layer is 256.

As before we test our proposed technique in two modes. In the first mode, we only use the given dataset. In the second one we augment this dataset with unsupervised data. The unsupervised data is obtained from the BCI competitions II, III and IV [63]. These datasets have different sampling rates, so all of them have be sub-sampled to 128 Hz. The same was done for the actual dataset [58] used in the experiments. Also the signals are normalized.

The results have been generated as before. The data is compressed to 25% and 50% of its original length and reconstructed using BSBL. The reconstructed signal is processed and classified using the techniques mentioned before. For our proposed techniques, the reconstruction and classification proceeds simultaneously. The reconstruction results are shown in Table IV and the classification results in Table V. The classification results are shown for 50% compression.

TABLE IV
EEG RECONSTRUCTION RESULTS

| Technique | 50% Compression | 25% Compression |
|---|---|---|
| BSBL | 0.112±0.062 | 0.240±0.084 |
| Prop. | 0.192±0.024 | 0.292±0.096 |
| Prop. Aug. | **0.060±0.006** | **0.096±0.012** |

TABLE V
EEG CLASSIFICATION RESULTS

| | Details | EMD | DTSTF | LPE | Prop. | Prop. Aug. |
|---|---|---|---|---|---|---|
| A | Eyes open | **88** | **88** | 86 | 86 | **88** |
| B | Eyes closed | 98 | 98 | 98 | 96 | **100** |
| C | Inter-ictal (epileptic focus) | 94 | **96** | 92 | 92 | **96** |
| D | Inter-ictal (Hipocam. region) | 95 | 95 | 92 | 93 | **96** |
| E | Ictal state | 92 | **94** | 90 | 92 | 92 |

From Table IV, we find that the reconstruction accuracy from our proposed technique is poor when we only use the test dataset [58]. This is because there is not enough data to learn the mapping; when we augment the dataset with unsupervised data the improvement is dramatic. It yields significantly better results than sparsity based methods. We see a similar speed improvement. BSBL takes about a minute to reconstruct 23.6 seconds of data whereas our proposed method takes only 1.8 seconds.

Table V shows the per class classification accuracy. Our proposed method (without augmentation) with only the supervised dataset does not yield very good results. This is likely to be an effect of overfitting of the autoencoder. With augmentation, the over-fitting issue is resolved and we get the best results overall results. It must be remembered that one cannot expect these results to match those in the published works; this is because the published papers use the groundtruth samples. Here the reconstructed samples are used. Owing to the reconstruction artifacts, the classification accuracy suffers.

## V. CONCLUSION

This work proposes a comprehensive solution for the tele health monitoring scenario. Prior studies addressed the problem in situ. Some studies concentrated on the acquisition and reconstruction of the signals whereas others focused on the analysis of these signals. During analysis it was assumed that the reconstruction is 'perfect'. This is not true. In prior studies [17-21] it has been shown that reconstruction artifacts do reduce the performance of automated analysis. This is mainly because prior techniques were based on hand-crafted feature extraction; they were dependent on the detection of peaks, troughs etc. Reconstruction artifacts corrupt these structures in the signal and hence the accuracy suffers.

There are several major contributions of this work. First, we propose a new approach for reconstruction. Prior compressed sensing based techniques are 'designed' assuming certain structures of the signal. In this work we 'learn' to reconstruct the signal; this does not require any assumption regarding the structure of the signal. As long as we have enough number of samples to train, our 'learned' approach excels over prior 'designed' techniques.

We employ an autoencoder for reconstruction. However, as mentioned before, reconstruction is not the final goal – signal analysis is. Here we introduce a linear map into the classifier that learns the class labels from the training samples. Thus our proposed label consistent autoencoder simultaneously learns to reconstruct and classify. We understand that learning such structure require sizable portion of the data; labeled data may not be always available. Our proposed label consistent autoencoder can work with both labeled and unlabeled data. If the data is labeled it learns to reconstruct and map (to class labels), if there is no class label associated with the sample, it only learns to reconstruct.

Usually Neural Networks are trained using some back-propagation (bp) algorithm. However our said architecture is non-linear and hence cannot be used solved using bp. We solve it using a recent class of optimization technique called Split Bregman.

The proposed semi-supervised stacked autoencoder is suitable for the said problem. However it can also be used when there is no necessity to reconstruct. One can input the same samples at the input and the output and the corresponding class labels (if available); this would learn an autoencoder based classifier which can be applicable to any problem. In the future we would test how the proposed method excels on benchmark deep learning datasets.

REFERENCES

[1] Z. Zhang, T. P. Jung, S. Makeig and B. D. Rao, "Compressed Sensing for Energy-Efficient Wireless Telemonitoring of Noninvasive Fetal ECG Via Block Sparse Bayesian Learning," in IEEE Transactions on Biomedical Engineering, vol. 60, no. 2, pp. 300-309, Feb. 2013
[2] J. Zhang, Z. Gu, Z. L. Yu and Y. Li, "Energy-Efficient ECG Compression on Wireless Biosensors via Minimal Coherence Sensing and Weighted \ell _1 Minimization Reconstruction," in IEEE Journal of Biomedical and Health Informatics, vol. 19, no. 2, pp. 520-528, March 2015.
[3] Z. Zhang, T. P. Jung, S. Makeig, Z. Pi and B. D. Rao, "Spatiotemporal Sparse Bayesian Learning With Applications to Compressed Sensing of Multichannel Physiological Signals," in IEEE Transactions on Neural Systems and Rehabilitation Engineering, vol. 22, no. 6, pp. 1186-1197, Nov. 2014.
[4] J. K. Pant and S. Krishnan, "Compressive Sensing of Electrocardiogram Signals by Promoting Sparsity on the Second-Order Difference and by Using Dictionary Learning," in IEEE Transactions on Biomedical Circuits and Systems, vol. 8, no. 2, pp. 293-302, April 2014.
[5] L. F. Polanía, R. E. Carrillo, M. Blanco-Velasco and K. E. Barner, "Exploiting Prior Knowledge in Compressed Sensing Wireless ECG Systems," in IEEE Journal of Biomedical and Health Informatics, vol. 19, no. 2, pp. 508-519, March 2015.
[6] Z. Zhang, T. P. Jung, S. Makeig and B. D. Rao, "Compressed Sensing of EEG for Wireless Telemonitoring With Low Energy Consumption and Inexpensive Hardware," in IEEE Transactions on Biomedical Engineering, vol. 60, no. 1, pp. 221-224, Jan. 2013.
[7] F. Pareschi, P. Albertini, G. Frattini, M. Mangia, R. Rovatti and G. Setti, "Hardware-Algorithms Co-Design and Implementation of an Analog-to-Information Converter for Biosignals Based on Compressed Sensing," in IEEE Transactions on Biomedical Circuits and Systems, vol. 10, no. 1, pp. 149-162, Feb. 2016.
[8] F. Chen, A. P. Chandrakasan, and V. M. Stojanovic, "Design and Analysis of a Hardware-Efficient Compressed Sensing Architecture for Data Compression in Wireless Sensors," IEEE Journal of Solid-State Circuits, Vol.47 (3), pp.744-756, 2012.
[9] N. Verma, A. Chandrakasan, "An ultra low energy 12-bit rate-resolution scalable SAR ADC for wireless sensor node", IEEE Journal of Solid State Circuits 42(6), 1196-1205, 2007
[10] R. Harrison and C. Charles, "A low-power low-noise CMOS amplifier for neural recording applications", IEEE Journal of Solid State Circuits, Vol. 38(6), 958-965, 2003.
[11] Z. Zhang and B. D. Rao, "Extension of SBL Algorithms for the Recovery of Block Sparse Signals With Intra-Block Correlation," in IEEE Transactions on Signal Processing, vol. 61, no. 8, pp. 2009-2015, April15, 2013.
[12] S. Aviyente, "Compressed Sensing Framework for EEG Compression", IEEE Workshop on Statistical Signal Processing, pp.181,184, 2007.
[13] S. Fauvel and R. K. Ward, "An Energy Efficient Compressed Sensing Framework for the Compression of Electroencephalogram Signals", Sensors 2014, 14(1), 1474-1496
[14] J. Chiang and R. K. Ward, "Energy-Efficient Data Reduction Techniques for Wireless Seizure Detection Systems", Sensors 2014, 14(2), 2036-2051
[15] M. Mohsina and A. Majumdar, "Gabor Based Analysis Prior Formulation For EEG Signal Reconstruction", Biomedical Signal Processing and Control.
[16] M. Hosseini Kamal, M. Shoaran, Y. Leblebici, A. Schmid and P. Vandergheynst. Compressive Multichannel Cortical Signal Recording. 38th International Conference on Acoustics, Speech, and Signal Processing (ICASSP), Vancouver, Canada, 2013.
[17] A. Majumdar, A. Gogna and R. Ward, "Low-rank Matrix Recovery Approach For Energy Efficient EEG Acquisition for Wireless Body Area Network", Sensors, Special Issue on State-of-the-art Sensor Technologies in Canada, Vol. 14(9), pp. 15729-15748, 2014.
[18] A. Majumdar and R. K. Ward, "Non-Convex Row-sparse MMV Analysis Prior Formulation For EEG Signal Reconstruction", Biomedical Signal Processing and Control, Vol. 13, pp. 142-147, 2014.
[19] A. Shukla and A. Majumdar, "Row-sparse Blind Compressed Sensing for Reconstructing Multi-channel EEG signals", Biomedical Signal Processing and Control, Vol. 18 (4), pp. 174–178, 2015
[20] A. Majumdar and R. K. Ward, "Energy Efficient EEG Sensing and Transmission for Wireless Body Area Networks: A Blind Compressed Sensing Approach", Biomedical Signal Processing and Control.
[21] A. Shukla and A. Majumdar, "Exploiting Inter-channel Correlation in EEG Signal Reconstruction", Biomedical Signal Processing and Control, Vol. 18 (4), pp. 49-55, 2015.
[22] P. C. Petrantonakis and Leontios J. Hadjileontiadis, "Emotion Recognition From EEG Using Higher Order Crossings", IEEE Transactions On Information Technology In Biomedicine, Vol. 14 (2), pp. 186-197, 2010.
[23] C. A. Frantzidis, C. Bratsas, M. A. Klados, E. Konstantinidis, C. D. Lithari, A. B. Vivas, C. L. Papadelis, E. Kaldoudi, C. Pappas and P. D. Bamidis, "On the Classification of Emotional Biosignals Evoked While Viewing Affective Pictures: An Integrated Data-Mining-Based Approach for Healthcare Applications", IEEE Transactions On Information Technology In Biomedicine, Vol. 14 (2), pp. 309-318, 2010.
[24] V. Bajaj and R. B. Pachori, "Classification of Seizure and Nonseizure EEG Signals Using Empirical Mode Decomposition", IEEE Transactions On Information Technology In Biomedicine, Vol. 16 (6), pp. 1135-1142, 2012.
[25] A. Temko, C. Nadeu, W. Marnane, G. B. Boylan and G. Lightbody, "EEG Signal Description with Spectral-Envelope-Based Speech Recognition Features for Detection of Neonatal Seizures", IEEE Transactions On Information Technology In Biomedicine, Vol. 15 (6), pp. 839-847, 2011.
[26] G. E. Hinton and R. R. Salakhutdinov, "Reducing the Dimensionality of Data with Neural Networks", Science. 2006 Jul 28;313(5786):504-7.
[27] P. Vincent, H. Larochelle, I. Lajoie, Y. Bengio and P.-A. Manzagol, "Stacked denoising autoencoders: Learning useful representations in a deep network with a local denoising criterion", Journal of Machine Learning Research, 11, 3371-3408, 2010.
[28] K. H. Cho, "Simple Sparsification Improves Sparse Denoising Autoencoders in Denoising Highly Noisy Images", ICML 2013.
[29] N. Wang and M. R. Lyu, "Extracting and Selecting Distinctive EEG Features for Efficient Epileptic Seizure Prediction," in IEEE Journal of Biomedical and Health Informatics, vol. 19, no. 5, pp. 1648-1659, Sept. 2015.
[30] M. Niknazar, S. R. Mousavi, B. Vosoughi Vahdat and M. Sayyah, "A New Framework Based on Recurrence Quantification Analysis for Epileptic Seizure Detection," in IEEE Journal of Biomedical and Health Informatics, vol. 17, no. 3, pp. 572-578, May 2013.
[31] Y. P. Lin et al., "EEG-Based Emotion Recognition in Music Listening," in IEEE Transactions on Biomedical Engineering, vol. 57, no. 7, pp. 1798-1806, July 2010.
[32] F. Riaz, A. Hassan, S. Rehman, I. K. Niazi and K. Dremstrup, "EMD-Based Temporal and Spectral Features for the Classification of EEG Signals Using Supervised Learning," in IEEE Transactions on Neural Systems and Rehabilitation Engineering, vol. 24, no. 1, pp. 28-35, Jan. 2016.
[33] M. Spüler, A. Walter, W. Rosenstiel and M. Bogdan, "Spatial Filtering Based on Canonical Correlation Analysis for Classification of Evoked or Event-Related Potentials in EEG Data," in IEEE Transactions on Neural Systems and Rehabilitation Engineering, vol. 22, no. 6, pp. 1097-1103, Nov. 2014.
[34] D. Menotti et al., "Deep Representations for Iris, Face, and Fingerprint Spoofing Detection," in IEEE Transactions on Information Forensics and Security, vol. 10, no. 4, pp. 864-879, April 2015.
[35] Li Deng, Jinyu Li, Jui, Ting Huang, Kaisheng Yao, Dong Yu, Frank Seide, Michael L. Seltzer, Geoff Zweig, "Recent Advances in Deep Learning for Speech Research at Microsoft", IEEE ICASSP 2013.
[36] Yoshua Bengio, Pascal Lamblin, Dan Popovici, Hugo Larochelle, "Greedy Layer-Wise Training of Deep Networks", NIPS 2007.
[37] Y. Bengio, "Learning deep architectures for AI", Foundations and Trends in Machine Learning, 2 (1),1-127. 2009.
[38] M. Chen, K. Weinberger, F. Sha, Y. Bengio, "Marginalized Denoising Autoencoders for Nonlinear Representation", ICML 2014.

[39] A. Makhani and B. Frey, "K-sparse Autoencoder", ICLR 2014.

[40] S Rifai, P Vincent, X Muller, X Glorot, Y Bengio, "Contractive auto-encoders: Explicit invariance during feature extraction", ICML 2011.

[41] J. A. Tropp and A. C. Gilbert, "Signal recovery from random measurements via Orthogonal Matching Pursuit", IEEE Transactions on Information Theory, 53 (12), 4655-4666, 2007.

[42] D. Donoho, Y. Tsaig, I. Drori and J.-L. Starck, "Sparse Solution of Underdetermined Systems of Linear Equations by Stagewise Orthogonal Matching Pursuit", IEEE Transactions on Information Theory, 58 (2), 1094-1121, 2012.

[43] D. Needell and J. A. Tropp, "CoSaMP: Iterative signal recovery from incomplete and inaccurate samples", Applied and Computational Harmonic Analysis, 26 (3), 301-321, 2009.

[44] I. Daubechies, M. Defrise and C. De Mol, "An iterative thresholding algorithm for linear inverse problems with a sparsity constraint", Communications on Pure and Applied Mathematics, 4 (57), 1413-1457, 2004.

[45] D. Donoho, "De-noising by soft-thresholding", IEEE Transactions on Information Theory, 41(3), 613-627, 1995.

[46] G. Cybenko, "Approximations by superpositions of sigmoidal functions", Mathematics of Control, Signals, and Systems, 2 (4), 303-314, 1989.

[47] K. Hornik, "Approximation Capabilities of Multilayer Feedforward Networks", Neural Networks, Vol. 4(2), pp. 251-257, 1991.

[48] Kolmogorov, A. N. (1957). On the representation of continuous functions of many variables by superposition of continuous functions of one variable and addition. Doklady Akademii Nauk SSSR, 144, 679-681; American Mathematical Society Translation, 28, 55-59 [1963].

[49] D. Donoho, "De-noising by soft-thresholding", IEEE Transactions on Information Theory, 41(3), 613-627, 1995.

[50] H. Larochelle and Y. Bengio, "Classification using Discriminative Restricted Boltzmann Machine", ICML 2008.

[51] Zhuolin Jiang, Zhe Lin, Larry S. Davis. "Label Consistent K-SVD: Learning A Discriminative Dictionary for Recognition". IEEE Transactions on Pattern Analysis and Machine Intelligence, 2013, 35(11): 2651-2664

[52] Rick Chartrand, "Nonconvex splitting for regularized low-rank + sparse decomposition", IEEE Trans. Signal Process., vol. 60, pp. 5810--5819, 2012.

[53] M. Llamedo, J.P. Martínez, "Heartbeat classification using feature selection driven by database generalization criteria", IEEE Transactions on Biomedical Engineering, 58 (3) (2011), pp. 616–625.

[54] Eduardo José da S. Luz, , Thiago M. Nunes, , Victor Hugo C. de Albuquerque, , João P. Papa, , David Menotti, "ECG arrhythmia classification based on optimum-path forest", Expert Systems with Applications, Vol. 40, Issue 9, July 2013, Pages 3561–3573

[55] R. J. Martis, U. R. Acharya and L. C. Min, "ECG beat classification using PCA, LDA, ICA and Discrete Wavelet Transform", Biomedical Signal Processing and Control, Vol. 8 (5), pp. 437-448, 2013.

[56] Jinkwon Kim, Hang Sik Shin, Kwangsoo Shin and Myoungho Lee, "Robust algorithm for arrhythmia classification in ECG using extreme learning machine", BioMedical Engineering OnLine, 8:31, 2009

[57] Ye, C., Coimbra, M. T., & Kumar, B. V. K. V. (2010). Arrhythmia detection and classification using morphological and dynamic features of ECG signals. In IEEE international conference on Engineering in Medicine and Biology Society (EMBC) (pp. 1918–1921)

[58] Time Series EEG Data: http://epileptologie-bonn.de/cms/upload/workgroup/lehnertz/eegdata.html. Last Accessed 9th May, 2016.

[59] K. Samiee, P. Kovács and M. Gabbouj, "Epileptic Seizure Classification of EEG Time-Series Using Rational Discrete Short-Time Fourier Transform," in IEEE Transactions on Biomedical Engineering, vol. 62, no. 2, pp. 541-552, Feb. 2015.

[60] R. B. Pachori and S. Patidar, "Epileptic seizure classification in EEG signals using second-order difference plot of intrinsic mode functions", Computer Methods and Programs in Biomedicine, Volume 113, Issue 2, February 2014, Pages 494–502

[61] V. Joshi, R. B. Pachori and A. Vijesh, "Classification of ictal and seizure-free EEG signals using fractional linear prediction", Biomedical Signal Processing and Control, Volume 9, January 2014, Pages 1–5.

[62] S. Altunay, Z. Telatarb and O. Erogul, "Epileptic EEG detection using the linear prediction error energy", Expert Systems with Applications, Volume 37, Issue 8, August 2010, Pages 5661–5665.

[63] BCI Compeition: http://www.bbci.de/competition/. Last accessed 9th May 2016.